\documentclass[10pt, conference, compsocconf]{IEEEtran}
\usepackage{cite}
\ifCLASSINFOpdf
   \usepackage[pdftex]{graphicx}
\else
\fi
\usepackage[cmex10]{amsmath}
\usepackage{algorithmic}
\usepackage[linesnumbered,ruled,vlined]{algorithm2e}
\usepackage[font=footnotesize]{subfig}

\begin{document}
\title{Building Halo Merger Trees from the Q Continuum Simulation}
\author{\IEEEauthorblockN{Esteban Rangel\IEEEauthorrefmark{1}\IEEEauthorrefmark{3},
		Nicholas Frontiere \IEEEauthorrefmark{2}\IEEEauthorrefmark{3}, 
		Salman Habib\IEEEauthorrefmark{3},
		Katrin Heitmann\IEEEauthorrefmark{3},\\ 
		Wei-keng Liao\IEEEauthorrefmark{1}, 
		Ankit Agrawal\IEEEauthorrefmark{1}, 
		Alok Choudhary\IEEEauthorrefmark{1}}
	\IEEEauthorblockA{\IEEEauthorrefmark{1}Electrical Engineering and Computer Science Department,
		Northwestern University\\
		Evanston, IL USA}
	\IEEEauthorblockA{\IEEEauthorrefmark{2}Department of Physics,
    University of Chicago\\
		Chicago, IL USA}
	\IEEEauthorblockA{\IEEEauthorrefmark{3}Argonne National Laboratory\\
		Lemont, IL USA}
}
\maketitle
\begin{abstract}
Cosmological N-body simulations rank among the most computationally intensive efforts today. 
A key challenge is the analysis of structure, substructure, and the merger history for many billions of compact particle clusters, called halos. 
Effectively representing the merging history of halos is essential for many galaxy formation models used to generate synthetic sky catalogs, an important application of modern cosmological simulations. 
Generating realistic mock catalogs requires computing the halo formation history from simulations with large volumes and billions of halos over many time steps, taking hundreds of terabytes of analysis data. 
We present fast parallel algorithms for producing halo merger trees and tracking halo substructure from a single-level, density-based clustering algorithm. 
Merger trees are created from analyzing the halo-particle membership function in adjacent snapshots, and substructure is identified by tracking the ``cores'' of merging halos -- sets of particles near the halo center. 
Core tracking is performed after creating merger trees and uses the relationships found during tree construction to associate substructures with hosts. 
The algorithms are implemented with MPI and evaluated on a Cray XK7 supercomputer using up to 16,384 processes on data from HACC, a modern cosmological simulation framework. 
We present results for creating merger trees from 101 analysis snapshots taken from the Q Continuum, a large volume, high mass resolution, cosmological simulation evolving half a trillion particles. 
\end{abstract}
\IEEEpeerreviewmaketitle
\section{Introduction}




The rich structure of the galaxy distribution in the universe results from gravity acting on smooth, almost featureless initial conditions. Due to the attractive gravitational instability, clumps of dark matter, called halos~\cite{diemand2011}\cite{knebe2011haloes}, form and merge in a hierarchical fashion as the universe expands and evolves. Halos host galaxies that assemble as ordinary (atomic) matter falls in, shock heats, cools, and forms stars; galaxies continue to evolve through a complex process of stellar evolution~\cite{mo2010galaxy}, astrophysical feedback mechanisms, and halo and galaxy mergers. For gravity-only cosmological simulations of dark matter, studying the galaxy distribution requires  a critical spatiotemporal analysis that is performed over a set of snapshots taken at time intervals short enough that the system dynamics are adequately captured~\cite{benson2012convergence}. The particle overlap between halos from every pair of adjacent snapshots is computed in order to identify the progenitors and descendant of each halo -- described using an interaction graph, called its merger tree~\cite{kauffmann1999}\cite{thomas2015sussing} (Figure~1). While approximate analytical models for agglomerative structure formation, such as the Extended Press-Schechter (EPS) formalism~\cite{bond1991excursion}, readily make use of a tree representation for describing hierarchical formation, it is much more difficult~\cite{srisawat2013sussing}\cite{wang2016sussing} with data from numerical simulations, where strict conformance to the tree representation breaks down due to the complex dynamical interactions of halos~\cite{bosch2016dissecting}. It is, therefore, a significant challenge for merger-tree building codes to deal with this issue in a way that robustly describes the growth and evolution of cosmic structure.

\begin{figure}[ht]
\begin{center}
\includegraphics[scale=.28]{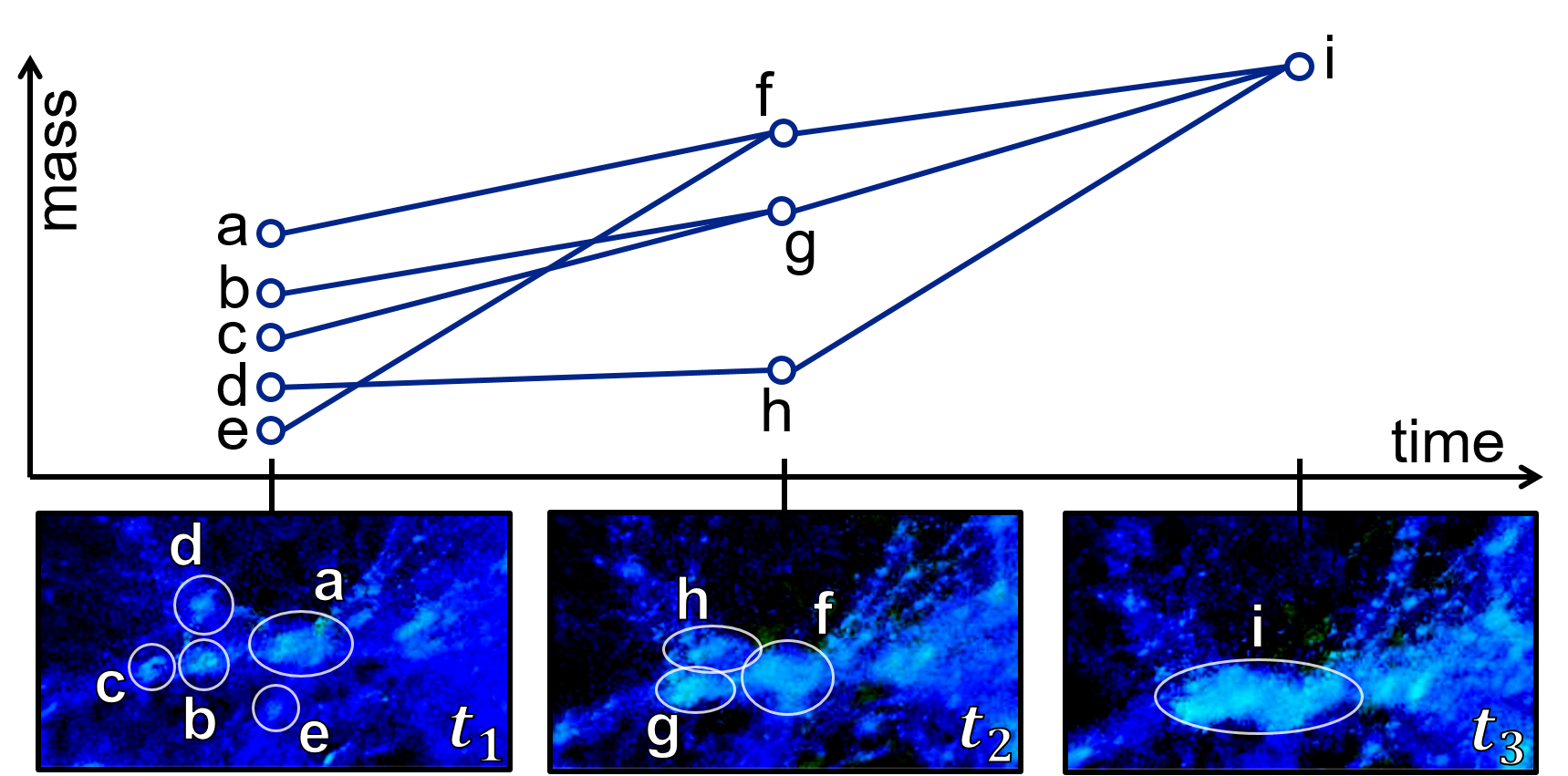}
\end{center}
\caption{The merger tree of a halo at time $t_3$ (upper) and the actual simulation particle rendering (lower), with halo regions found by the halo finding step superimposed~\cite{preston2016integrated}.}
\label{halo_explanation}
\end{figure}

In the modern cosmological context, halo merger trees form the basis for implementing galaxy formation frameworks that either follow empirical methods~\cite{hearin2013sham} or semi-analytic modeling~\cite{benson2012galacticus}. 
Parameters are tuned and validation carried out by confronting these results with observational data.
The current state-of-the-art in N-body codes consists of tracking halos using clustering or density-based algorithms, which use spatial information encoded in the particle positions, and then adding phase space correlations, by including the particle velocities. Multi-step merger tree histories are also used to add supplementary information. Hierarchical merger schemes track sub-halos within halos, and sub-halos within sub-halos. Because sub-halos can be disrupted by gravitational tidal forces within halos, whereas galaxies mostly survive, it is not simple to identify where the associated galaxies might reside in a halo. We address this last problem by following dense halo `cores' as described below.

We present an approach for constructing merger trees strictly from the particle membership function generated by a percolation-based, FOF (`friends of friends')~\cite{davis1985evolution} halo finder. Our algorithm makes no assumptions about the underlying physics of the simulation and solely performs set intersections on the particle IDs when comparing halos in adjacent snapshots to properly follow their evolution. While this approach is conceptually straightforward, accuracy requirements dictate significantly higher memory and computation costs compared to approximate strategies for matching halos based only on their aggregate statistics (these methods would fail the accuracy requirements).
As such, we propose a distributed memory parallelization strategy for multiple set intersection -- based on array intersection of the particle IDs -- that uses the same uniform spatial volume decomposition as the underlying simulation. 
We found that traversing the snapshots in temporally reverse order allows robust identification of splitting halos -- a major source of inconsistencies in merger tree construction -- and we also propose a method to augment the halo catalogs and alter the membership arrays accordingly. The key contributions of this work are:
\begin{enumerate}
\item Development of a novel parallel algorithm for robustly tracking halo formation and substructure for galaxy formation models that does not require complex subhalo definitions.
\item A memory efficient implementation allowing for analysis on reduced computing resources scalable to the largest state-of-the-art simulations. 
\end{enumerate}

All of our experiments were performed on Titan, a Cray XK7 supercomputer at the Oak Ridge Leadership Computing Facility (OLCF). 
Our results were gathered from analyses performed on N-body (gravity only) simulations generated by the HACC (Hybrid/Hardware Accelerated Cosmology Code) framework~\cite{habib2016hacc}. 
The halo data used as input for all results presented were generated by the FOF halo~\cite{fasel2011cosmology} and core finder integrated into HACC's analysis framework. 
We used halo and core data from a moderately sized (1 billion particle) simulation to demonstrate the scaling performance at varying times in the clustered regime of a cosmological simulation. 
Our experiments show a speedup of up to $\sim$71$\times$ when scaling to 256 processes. 
To demonstrate the performance on a modern (extreme-scale) simulation, we ran our code on the largest complete simulation data set ever produced with a particle mass resolution of $m_p \simeq 1.5\cdot10^8 M_{\odot}$. 
The underlying HACC simulation, Q Continuum~\cite{heitmann2015q}, evolves over half a trillion particles -- approximately half of which make up nearly 400 million halos per snapshot -- and can be used for the construction of detailed synthetic sky catalogs, where an accurate formation history of the halos -- represented by the merger trees -- provides a critical foundation for different modeling methodologies.

\section{Background}

The burgeoning increase in current cosmological knowledge is a direct result of the remarkable success of wide and deep multi-wavelength observations of the sky. The next generation of surveys will be limited not by finite statistics, but by low levels of systematic errors, both observational and modeling-related. Because cosmology is an observational and not an experimental science, the resulting emphasis on improved and robust 
modeling of large-scale sky survey observations is a key driver for the continuing development of high resolution, large-volume, cosmological simulations. In the case of optical surveys, the aim is to construct synthetic sky catalogs that place different types of model galaxies in the appropriate physical locations. A necessary requirement for doing this by any of the current methods is the extraction of a detailed halo evolution history from the underlying high-resolution simulation.

\subsection{Dark Matter Halos}

In the currently standard cold dark matter (CDM) cosmology, the initial density perturbations are specified by a Gaussian random field with a scale-dependent power spectrum. As time evolves, all scales increase in amplitude due to the linear gravitational instability. Small scales collapse first and the collapsed objects continue to merge, to yield a variety of halos with different masses, profiles, and substructure. The halo distribution continues to evolve to the current epoch.

In numerical simulations, halo finding is a first level analysis task that is performed on the raw simulation data; for particle-based simulations, halos are collections of particles identified by some means of computing the local density. Because the mass resolution of an N-body simulation (smallest mass that can be tracked) is necessarily finite, only halos above a certain threshold mass can be reliably identified. At late times, about half the particles in simulations can be found in halos. 

Halo identification can be performed by a number of methods; the two most popular are spherical overdensity (based on an average density criterion) and FOF (based on an approximate isodensity criterion). The two definitions are in principle not the same but because the profiles of halos~\cite{navarro1997universal} tend to follow a universal form (NFW, after Navarro-Frenk-White), there is a tight correlation between the two~\cite{lukic2009structure}.

It should be noted that subhalos do not form within halos but result from a smaller mass halo merging with a much larger object. Given the finite halo mass threshold in a simulation, every subhalo is therefore, at some point, an isolated halo with no substructure. There are a number of robustness issues with identifying and tracking subhalos in simulations, especially when the numbers of associated particles are small $(\sim 20)$ and near the central high-density region of the host halo. A large dynamic range in mass is needed for making sure that halo substructure is being properly treated. In the case of the Q Continuum run, a typical galaxy group or galaxy cluster scale halo ($\sim 10^{13-15}M_\odot$) is resolved by $\sim 10^{5-7}$ particles, which easily meets the requirements.

\subsection{Halo Merger Trees}
Merger tree codes link together halos (and subhalos) from temporally adjacent simulation snapshots where halo finding was performed. The goal is to accurately represent the hierarchical formation of dark matter halos. The physical context relates to modeling galaxy formation within halos; in principle this can be performed in a history-independent manner using the halo occupation distribution (HOD) method, which populates a halo with galaxies based only on mass. More sophisticated approaches, however, need to incorporate the merger history as well as knowledge of the halo environment. It should be kept in mind, however, that halo formation and evolution is a complex dynamical process -- the assumption that a simple merger tree picture is the correct spatio-temporal description is only an idealization.

\subsection{Galaxy Positions within Halos}

In the conventional approach, galaxy positions are associated with subhalo positions. In the situation where there are none or very few subhalos in a main halo, this procedure is reasonable, however, in large halos that can contain hundreds of galaxies, this sort of procedure can run into trouble. Subhalos tend to reside in the outer parts of such halos, since tidal disruption reduces their population at smaller radii. It is known that galaxy populations do not have this property and continue to rise in density as the radius decreases (although not with the same slope as the central dark matter distribution itself).

The reason for this behavior is that galaxies are more compact and robust than their host halos -- they can survive tidal forces that the subhalos cannot; those that exist having had most of their dark matter halo stripped away are termed `orphan' galaxies. In order to track orphan galaxies along with the more usual subhalo approach we use a method based on tracking the high-density cores of halos, by identifying a central bundle of particle trajectories associated with each halo and then following this bundle until the end of the simulation. The advantage of this method is that it identifies subhalos as well as small-scale overdense regions which are potential sites for orphan galaxies; the merger and diffusion of the trajectory bundles mimic galaxy mergers in the halo and the formation of intracluster light -- a luminous component that has been stripped from cluster galaxies. 

The assumption that the observed galaxy distribution can be obtained following the core positions (along with a few modeling assumptions) can be tested by comparison against observations from surveys such as the Sloan Digital Sky Survey (SDSS); initial results show excellent agreement between the model predictions and the observed galaxy distribution profiles.

\section{Related Work}\label{related_work}

Analytical models for halo formation, such as EPS~\cite{bond1991excursion}, do not account for halo splitting, however, it is a robustly observed phenomenon in numerical N-body simulations and accounts for $\sim$5\% of all halos we observed in our analyses. 
The EPS model describes a purely hierarchical assembly of halos through mergers and is the basis for the widely accepted merger tree representation~\cite{thomas2015sussing} that has become the standard format for describing halo formation in simulations.
The shortcomings of the overly simplified tree representation has given rise to many complicated algorithms~\cite{srisawat2013sussing} for producing merger trees aiming to provide physically consistent properties (mass, position, velocity), some of which even alter the input halo catalog by adding missing (or removing spurious) objects in order to achieve consistency~\cite{behroozi2012gravitationally}.

\begin{figure}[tbp]
\begin{center}
\includegraphics[scale=.3]{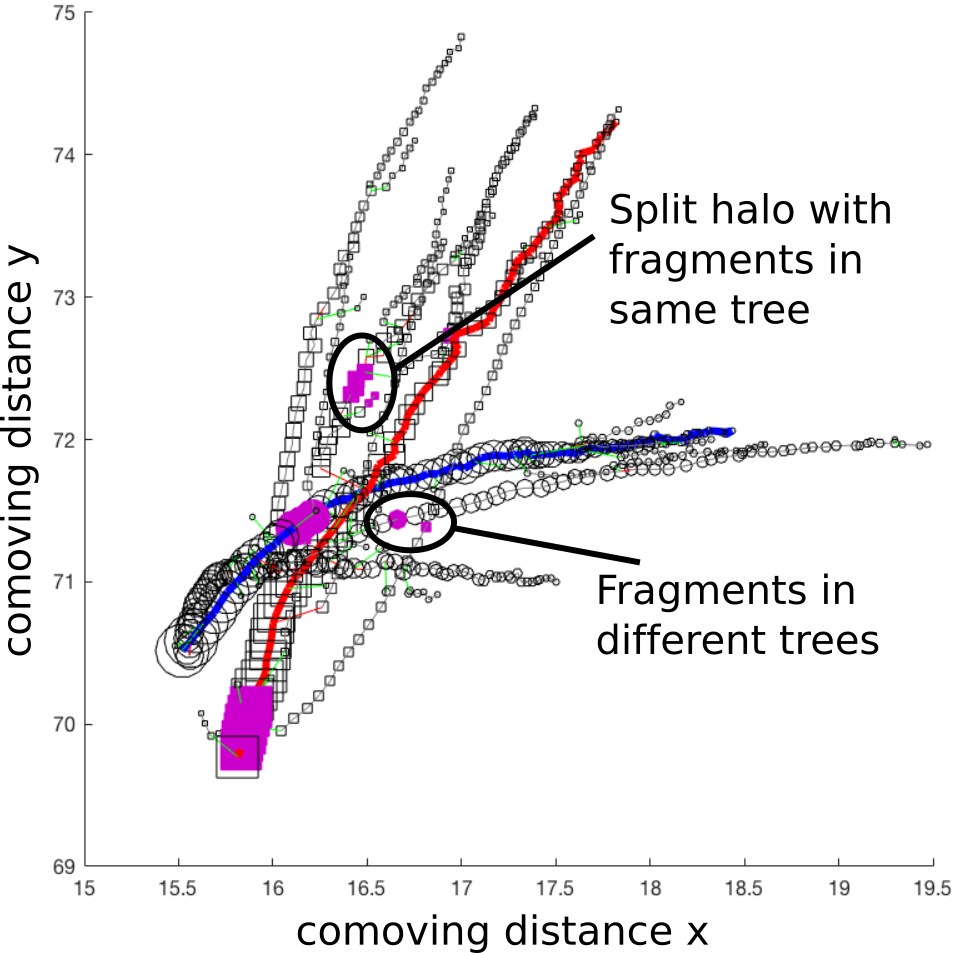}
\end{center}
\caption{Two merger trees created by our splitting algorithm for non-hierarchical FOF halos with crossing trajectories. Fragments (solid purple) from a splitting halo finally merge with separate halos in the final snapshot. Plot markers of the same type indicate progenitors of the same final halo. The main halo path (backbone) for the two trees are in bold and colored red and blue respectively. Marker size is proportional to FOF halo mass and is scaled for visualization purposes. Each marker represents a halo from a distinct analysis snapshot. The mass of fragment halos is calculated using our approximation method. Distance units are in Mpc h$^{-1}$, one axis is suppressed.}
\label{fragment_diff_trees_an}
\end{figure}

Many tree building codes create merger trees from subhalos~\cite{behroozi2012rockstar}, essentially creating a forest of subhalo trees for each final halo in the last analysis snapshot.
These forests of trees can be difficult to interpret and can even become entangled when subhalos leave their hierarchical halo group.
When a subhalo leaves the hierarchical group to which it was previously associated, the halo group is referred to as a \textit{splitting} halo, and the resulting split groups as \textit{fragment} halos. 
For some types of analysis, e.g., calculating merger rates, trees from the halo groups (complete halos) are required,
leading to several approaches~\cite{genel2009halo} to explicitly handle halo splitting.
The approaches broadly fall into the following few categories: snipping, stitching, and splitting, and are typically a post-processing step on the subhalo trees themselves in an effort to essentially disentangle the forests. 

The simplest of these methods is snipping~\cite{fakhouri2008nearly} where fragments are not allowed to have progenitors. 
In addition to mergers with a fragment being potentially over counted (splitting and re-merging can occur several times) the entire history of the fragment can be wrongly associated with the tree of the first merger if the splitting halo finally merges with different halos. (See Figure \ref{fragment_diff_trees_an}.)
The stitching method improves on snipping by merging fragments with the halo of a future merger, basically closing the cycle in a graph to conform to a tree representation.
This method also suffers from inaccuracies when fragments end up merging with different halos in the final analysis step.
Splitting is the most effective at producing trees with consistent properties. 
Any time a split is observed the fragments are considered halos just as in the original set. 
Splitting only considers the final merger as the true merger and does not produce remnant fragments. 

\section{Merger Trees}\label{mt}
Our method builds halo merger trees from non-hierarchical FOF halos; that is, input halo catalogs need not contain subhalo information. 
As such, we define a ``split'' event not as the loss of a subhalo as described in Section~\ref{related_work}, but rather, as the loss of a significant number of particles to another FOF halo which is not the direct descendant. 
In this way, our approach does not rely on the accuracy or robustness of a subhalo finder's ability to identify subhalos.

\subsection{Design}\label{mt_design}
Our merger tree algorithm determines the evolution of halos by comparing the halo-particle membership function from adjacent snapshots.
The membership function is a simple 2-column array that acts as a map between the particle IDs and their host halo TAG; 
simulation particles not in any halo are omitted. 
The algorithm iterates over pairs of adjacent snapshots in a temporally reversed order, i.e., the last simulation snapshot generated is the first to be processed, and connects halos based on the overlap in their particle ID sets.
Traversing the snapshots in reverse greatly simplifies the ability to identify and correct for the temporary merging of halos, the failure of which can lead to large mass fluctuations in the accretion history that are unphysical, potentially having disastrous consequences on downstream analysis. 
A key advantage of this strategy is that it is effective at fully splitting temporarily merged halos across any number of snapshots with only two analysis snapshots in memory at any given time.

\subsubsection{Construct Halo-Halo Intersection Matrix}
The central operation in the algorithm is computing the cardinality of particle set intersections on the Cartesian product of halos taken from adjacent snapshots.
Formally,
\begin{equation}
 \left \vert{h_a \cap h_b} \right \vert \ ,\ \forall h_a \in H_A \ \textrm{and} \ \forall h_b \in H_B
\end{equation}
where $H_A$ and $H_B$ are sets of halos from adjacent analysis snapshots, and snapshot $A$ comes before $B$.
Conceptually, the results are stored in an $m \times n$ matrix, $\mathbf{C}$, where $m=\vert H_B \vert$ and $n=\vert H_A \vert$.
Hence, $\mathbf{C}_{i,j} = \left \vert{h_b^i \cap h_a^j} \right \vert$. 
The choice for this representation where the older halos are represented as columns is motivated by our implementation and for simplicity we use this convention throughout the paper, although conceptually it is an arbitrary decision.
\\
\subsubsection{Identify Split Events}
Once the matrix, $\mathbf{C}$, has been completely constructed, it is analyzed to determine splitting events by examining the columns of the matrix.
Each column represents the intersection of particles from an older halo into the set of newer halos. 
A parameter of the algorithm, $min\_split$, defines the minimum number of particles needed for any intersection to be deemed significant to the analysis; 
typically this is chosen close to the minimum halo size.
Splitting halos in $H_A$ are simply identified by checking if the corresponding column in $\mathbf{C}$ has no more than one entry greater than $min\_split$.
In the case where a split halo is identified, a new halo for each split component is inserted into the catalog and the splitting halo is removed. 
We must also alter the original membership function to reflect the insertions and deletion by reassigning the particles to their corresponding new halos.
Finally, properties of the new halos, namely position, velocity, and mass, must be calculated. 
The position and velocity are extrapolated from the descendant halo, however, new halo masses are proportionally assigned from the original halo mass according to a split fraction that is calculated as, 
\begin{equation}
	\frac{\mathbf{C}_{i,j}}
	 {\sum_{j=1}^{n} \left\{
\begin{array}{ll}
      0 & \mathbf{C}_{i,j} < min\_split \\
      \mathbf{C}_{i,j} & \mathbf{C}_{i,j} \geq min\_split \\
\end{array} 
\right.}
\end{equation}
which ensures mass fluctuations are smoothed, but still driven by the dynamics of the simulation.
\\
\subsubsection{Update Descendants}
Descendant halos are determined by using a standard merit function. 
In our approach, $h_b$ is the descendant of $h_a$ when the halo pair have the greatest fractional overlap.
$\forall h_a^j \in H_A$ 
\begin{equation}
descendant(h_a^j) = \underset{h_b^i \in H_B}{\arg\max} \left ( \mathbf{C}_{i,j} \right )
\end{equation}

\subsection{Implementation}\label{mt_imp}
 The domain of the underlying simulation is decomposed into equal 3D sub-volumes for parallelization. The sub-volume size is determined using domain knowledge, such that the maximum distance a particle can travel during the simulation is not beyond an adjacent sub volume. Halos are typically found using the same decomposition, and we leverage this in building merger trees by pruning the descendant search space, thereby limiting halo particle overlap calculations to neighboring sub-volumes. That is, for each sub-volume at time, $t-1$, a process must search the local -- and all 26 adjacent sub-volumes at time $t$ to complete the iteration. The next iteration for incrementally building trees finds descendants for halos at time $t-2$, in the $t-1$ snapshot. There are no interprocess data dependencies when finding descendants for halos during each iteration, however, there are dependencies between iterations, since the data at time $t-1$ may be altered by our splitting strategy. 

During an iteration, each of the $N$ sub-volumes can be considered an independent sub-problem that requires data from sub-volume, $i$, at time $t-1$, and data from sub-volume, $i$, at time $t$, including all adjacent sub-volumes of, $i$, at time $t$. Difficulty arises due to memory requirements -- typically holding all 26 neighboring sub-volumes is beyond the physical available memory, but all data is not needed at once, and thus can be iteratively processed reusing available memory. Additionally, the halo-halo intersection matrix is largely sparse, with most of the columns having at least one non-zero row. As such, we implement the 2D matrix as an array of lists, where the columns are static and rows are dynamically created; missing rows assumed to be zero. 

The algorithm described in section \ref{mt_design} is implemented in C++ using MPI for portability across parallel systems. 
We break down the algorithm into a set of procedures and provide a complexity analysis for each as they are presented, providing pseudocode where necessary. 
The algorithmic complexity is described using the number of halos, $N_h$, and the total number of particles in all halos, $N_p$, where $N_h \ll N_p$ (typically by $\sim10^3 \times$). 
The procedures called during each iteration are as follows:
\begin{enumerate}
\item{\texttt{ReadSnapshot}} -- reads data from sub-volumes of the simulation snapshot and (if necessary) spatially redistributes the halos and member particle IDs based on the position of the halo into a new uniform spatial decomposition. 
For convenience, the halo-particle membership function is also sorted by particle ID in this step. 
Requires all-to-all communication for spatial redistribution of the  halos and membership function. 
$\mathcal{O}(N_p\ log_2\ N_p)$
\item{\texttt{PrepSnapshot}} -- builds a map from the halo IDs (64 bit integers) to the corresponding position in the array of halos. 
Also performs consistency verification guaranteeing halo IDs are not duplicated and all halo IDs in the membership function are present. 
$\mathcal{O}(N_h\ log_2\ N_h+N_p\ log_2\ N_h)$
\item{\texttt{RemoveDuplicatePIDs}} -- removes duplicate particle IDs within the same halo-particle membership function. 
Imperfections in halo finders will sometimes produce duplicate, or partially duplicated halos. 
Requires neighbor-to-neighbor synchronous exchange of the membership function. 
$\mathcal{O}(N_p\ log_2\ N_h)$
\item{\texttt{FillMergerTable}} -- computes the cardinality of particle ID intersections on the Cartesian product of halos, storing the result in a sparse matrix. 
Time complexity is $\mathcal{O}(N_p\ log_2\ N_h)$. See Algorithm \ref{FillMergerTable} for pseudocode.
\item{\texttt{AugmentTable}} -- augments the sparse matrix of intersection cardinality with information from spatially adjacent sub-volumes.
Requires a neighbor-to-neighbor synchronous exchange of the partial (only those not found locally) membership function.
$\mathcal{O}(N_p\ log_2\ N_h)$
\item{\texttt{AnalyzeTable}} -- identifies splitting halos and alters the halo catalog by inserting and deleting. Sets the descendants according to the merit function.
$\mathcal{O}(N_h)$
\item{\texttt{ChangeSplitTags}} -- alters the halo-particle membership function for halos that are identified as splitting. 
Requires a neighbor-to-neighbor synchronous exchange of the partial membership function.
$\mathcal{O}(N_p)$
\item{\texttt{SwapSnapshot}} -- copies memory from snapshot $B$ to memory buffers for snapshot $A$ so that the next snapshot is read into $B$.
$\mathcal{O}(N_h+N_p)$
\\
\end{enumerate}
The memory requirement for executing the algorithm is largely dominated by the halo-particle membership function. We require two snapshots for computing particle overlap and additionally allocate another (snapshot sized) buffer for redistribution and particle exchanges. Each particle in the membership function requires two 64-bit integers (16 bytes), halos require 90 bytes (including halo properties), and each sparse table entry requires 32 bytes. Since the size of the halo-halo intersection matrix (in bytes) is set by the number of non-null intersections, it is proportional to the number of halos in the snapshots compared. 

The input halo data is stored on disk and organized by data blocks that correspond to spatially continuous sub-volumes of the simulation. 
The decomposition in the files is not preferred for offline processing when using hardware where there is available a greater amount of memory per process. 
In such a case, we redistribute the halos according to the center found by the halo finder.
Otherwise, when the number of MPI processes is commensurate with the number of blocks, they are simply read by the process matching the block index. 
In both cases, logically mapping the uniform volume decomposition to the processes is accomplished using the Cartesian virtual topology functions of MPI.
\\
\begin{algorithm}
$cell \leftarrow arrData[col]$\;
\While{$cell$ != $NULL$}{
\If{$cell.row==row$}{
$cell.count++$\;
\Return
}
\Else {
  $cell \leftarrow cell.next$\;
}
}
$cell \leftarrow arrData[col].append(row)$\;
$cell.count++$\;
\Return
\caption{SparseArray.Increment($row,col$)}
\label{FillMergerTable}
\end{algorithm}

\begin{algorithm}[h!]
$i \leftarrow 0$\;
$j \leftarrow 0$\;
$n \leftarrow H_A.membership$ length\;
$m \leftarrow H_B.membership$ length\;
$SparseArray.Initialize(|H_A|)$\;
\While{$i < n$ and $j < m$}{
$p \leftarrow H_B.membership[j]$\;
$q \leftarrow H_A.membership[i]$\;
\If{$p$.pid == $q$.pid}{
$row \leftarrow H_A.map.find(p.tag).value$\;
$col \leftarrow H_B.map.find(q.tag).value$\;
$SparseArray.Increment(row,col)$\;
}
\ElseIf{$p.pid < q.pid$}{
Insert $p$ into $UnmatchedPIDs$\;
$i++$\;
}
\Else{
$j++$\;
}
}
\Return{$SparseArray,UnmatchedPIDs$}
\caption{FillMergerTable($H_A,H_B$)}
\label{FillMergerTable}
\end{algorithm}
The \texttt{AugmentTable} procedure is similar to \texttt{FillMergerTable}, with the substitution of the input $H_B$ with $UnmatchedPIDs$ of each neighbor.  
After the \texttt{AnalyzeTable} procedure identifies splitting halos by scanning the columns of the halo-halo intersection matrix, \texttt{ChangeSplitTags} renames the halo TAGs associated with particle IDs mapped to cells marked as split fragments. Deciding which particles must be changed is performed similarly to filling the matrix and therefore also requires neighbor data exchange. 

\section{Tracking Halo Substructure}\label{core_tracking}
Our method for substructure tracking builds on the premise that subhalos do not form inside a host halo. 
The motivation for this approach comes from the  theory of hierarchical structure formation. 
Nearly all observed substructure is the result of merging with other halos, in other words, no excess subhalo clustering occurs inside the host halo and we can therefore leverage the merger history and core information to understand halo substructure evolution.
Stated as the set of $N_{core}$ particles nearest in distance to the Most Bound Particle (MBP) center, a halo core is a robust estimator for not only the position and velocity, but also the stability of a halo once it merges and becomes substructure of the host. 
The MBP is formally defined as 
\begin{equation}
\text{arg}\,\min\limits_{i \in H}\ \left ( - \sum_{j \in H, i \neq j} \frac{m_j}{d_{i,j}}   \right )
\end{equation}
where $m_j$ is the particle mass and $d_{i,j}$ is the Euclidean distance between particles. 


The intended application for tracking the cores of merging halos is to act as a proxy for potential galaxies; more commonly done by identifying subhalos, a process that itself can have issues with robustness~\cite{bosch2016dissecting}. 
By tracking a set of particles at each step, additional statistical information about the core, e.g., a description of compactness, can be used to infer galaxy mergers and disruptions, a task that would traditionally not only require robustly identifying subhalos, but also building the corresponding subhalo merger trees.

\subsection{Design}\label{core_desigh}
\begin{figure}[tbp]
\begin{center}
\includegraphics[scale=.28]{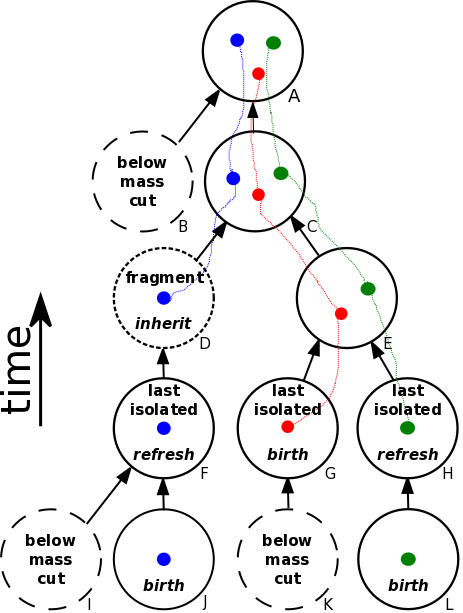}
\end{center}
\caption{An example of how our core tracking algorithm identifies substructure within halos by utilizing merger tree information. Long dash lines indicate halos below a mass threshold where our code does not consider a core to exist. If a halo is isolated, i.e., has no previous mergers, the core is ``refreshed'' and the set of particles nearest to the halo center is rediscovered. Fragment halos inherit the core from the main progenitor. The process is forward in time.}
\label{core_merger_tree}
\end{figure}

Our core tracking algorithm is conceptually straightforward and is only complicated by fragmented halos introduced during merger tree construction. 
Since cores are found during the halo finding phase (prior to building merger trees), fragment halos will not have halo core information.
The solution we propose is to have any fragment halo inherit the set of core particle IDs from the main progenitor, see Figure~\ref{core_merger_tree}. 
In the case of a long lasting temporary merger where several fragment halos occur in succession, this process occurs recursively. 
We have observed this strategy to work well when recovering cores for fragment halos, seeing only 0.5\% of all tree leafs to be fragment halos.

\subsubsection{Assigning Cores to Halos}
Cores are assigned to halos using the merger tree information, specifically, the merger of an isolated halo (itself not the product of mergers) passes its current core to the descendant. 
Unlike an isolated halo where the nearest $N_{core}$ particles are rediscovered, the particles making up the cores of non-isolated halos never change.
Non-isolated halos pass along their core set to their descendants, essentially aggregating cores from all progenitors. 

\subsubsection{Update Core Positions}
At each analysis step, the set of particles that were once flagged by the halo finder must be used to update the positions of particles in non-isolated halo cores. 
The core properties (position, velocity, compactness) are then recomputed from their updated particles. 

\section{Results}~\label{results}
To demonstrate the effectiveness of our method we present results for building merger trees with core tracking on two cosmological simulations, the Q Continuum simulation and the AlphaQ simulation carried out with the HACC framework (Hardware/Hybrid Accelerated Cosmology Code)~\cite{habib2016hacc}. Both simulations use the same cosmological parameters, given by the dark matter and baryonic content of the Universe ($\omega_{\rm cdm}=0.1109$ and $\omega_b=0.0226$), the normalization and slope of the linear power spectrum ($\sigma_8=0.8$ and $n_s=0.963$), the dark energy equation of state, $w=-1$, and the Hubble parameter $H_0=71.0$. This cosmological model is close to the measurements obtained by the Wilkinson Microwave Anisotropy Probe (WMAP) during its seventh year operation~\cite{komatsu2011seven}. The QContinuum simulation covers a volume of (1300Mpc)$^3$ and evolves 8192$^3$ particles, leading to a particle mass of $m_p\sim 1.5\cdot 10^8$M$_\odot$. It is one of the largest cosmological simulations at this resolution ever carried out. It was run on Titan, a Cray XK7 supercomputer at the Oak Ridge Leadership Computing Facility (OLCF), using more than 90\% of the machine. The simulation generated more than 2PB of raw data. Details about Q Continuum simulation including HACC's implementation on GPUs can be found in Ref.~\cite{heitmann2015q}. The AlphaQ simulation is a smaller test simulation that is mainly used to develop new analysis methods. It covers a volume of  (360.5Mpc)$^3$ and evolves 1024$^3$ particles, leading to a mass resolution of $3.2\cdot 10^9$M$_\odot$. The AlphaQ simulation was carried out on Cooley, a GPU accelerated cluster at the Argonne Leadership Computing Facility.

\subsection{Data}
The simulation codebase has an integrated analysis framework and includes the following tools used to generate our test data: a parallel halo finder~\cite{fasel2011cosmology} that implements the FOF algorithm, a halo center finder for finding the Most Bound Particle (MBP), and a core finder to identify simulation particles near the MBP center of each halo. 
HACC's analysis tools were used to generate the halo and core data used as in our experiments. 
All data is stored and read using GenericIO, a library for writing self-describing scientific data files that is optimized for large-scale parallel file systems.
The simulation and analysis tools partition the domain into equally sized sub-volumes which are then stored as contiguous blocks of data and mapped to the decomposition. 
The FOF halo data for creating merger trees from the Q Continuum simulation was generated from 101 simulation snapshots totaling nearly 200TB, most of which are the membership functions mapping particles to halos. 
The AlphaQ simulation halo analysis is also performed on the same number of snapshots and is approximately 500GB of data in total. 

\subsection{Environment}
Our experiments were carried out on Titan, a Cray XK7 supercomputer located at Oak Ridge Leadership Computing Facility (OLCF).
We run with a configuration of 16 MPI ranks/node (1 per physical core) unless we are limited by the amount of physical memory available.
On Titan, each compute node contains a single 16-core processor and 32GB of RAM.

\subsection{Experiments}
To show how our merger tree algorithm performs with data from a real cosmological simulation, we chose the moderately sized (1 billion) particle simulation called AlphaQ. 
\begin{figure*}[h!]
\centering
\subfloat[]{\includegraphics[scale=.2]{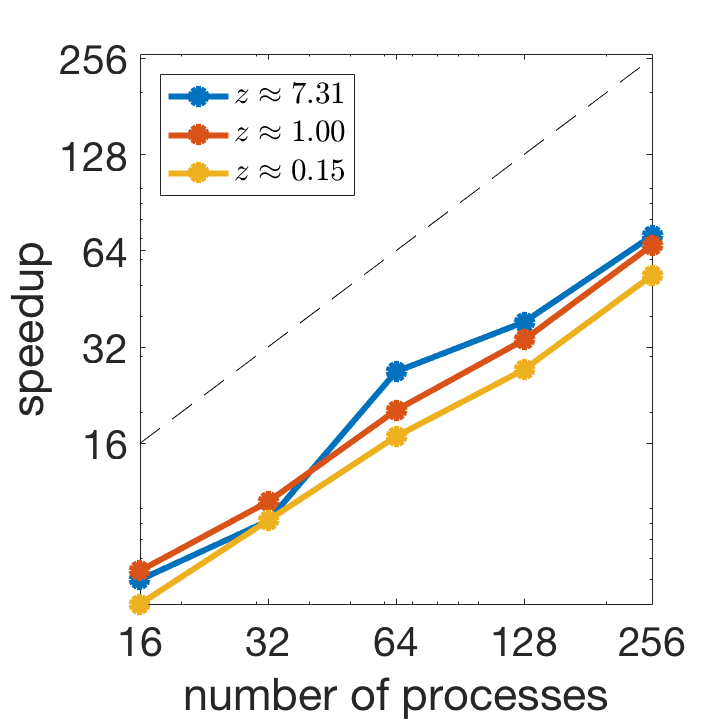}\label{speedup}}
\subfloat[]{\includegraphics[scale=.2]{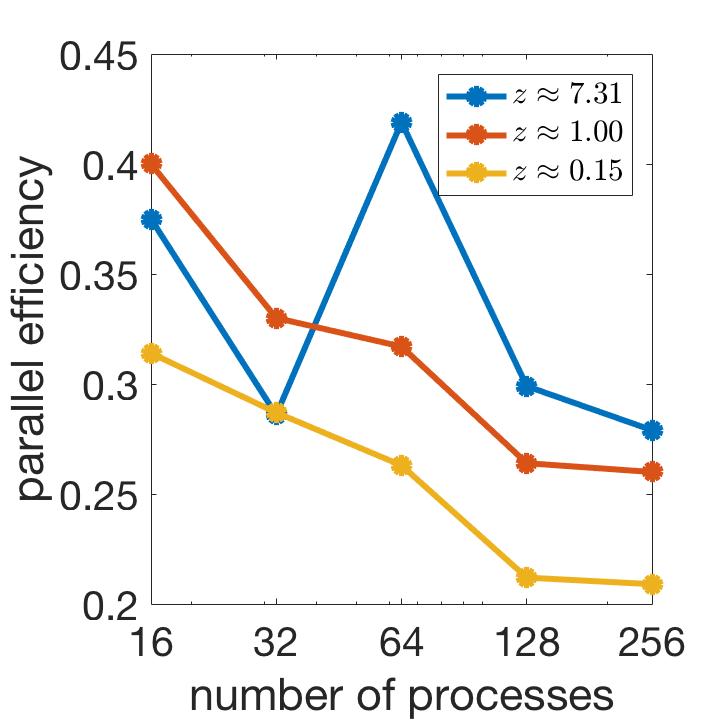}\label{efficiency}}
\subfloat[]{\includegraphics[scale=.2]{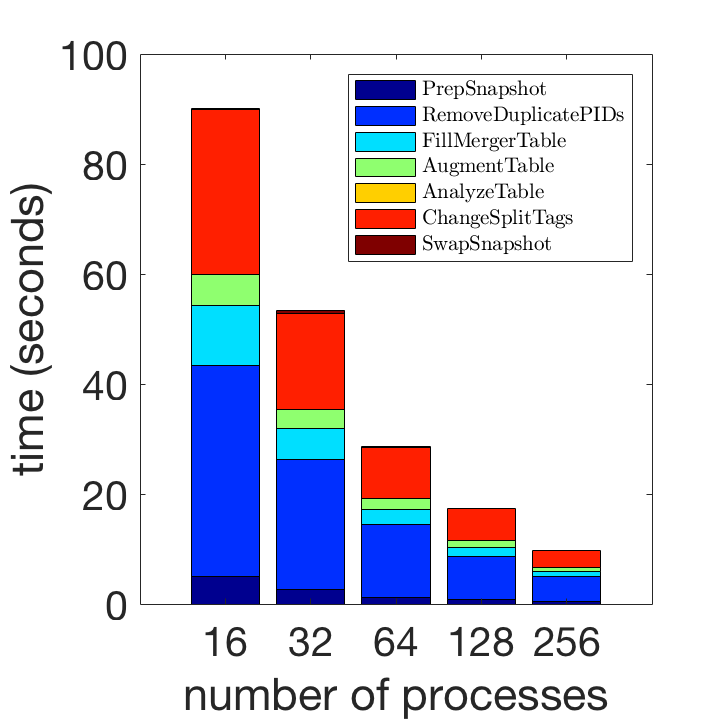}\label{breakdown}}
\caption{Speedup \ref{speedup} and parallel efficiency \ref{efficiency} of overall performance of the merger tree building algorithm on the AlphaQ simulation at selected snapshots, labeled as $z$ for redshift, used to describe cosmic time. $z=0$ indicates now, or the final timestep of a cosmological simulation, and larger values indicate earlier time. Generally, higher $z$ indicates more homogeneity in the mass distribution and lower values of $z$ indicate more clustering, and as a result greater load imbalance. \ref{breakdown} shows a breakdown of the individual phases of computation at $z \approx 1.0$ for the results reported in \ref{speedup} and \ref{efficiency}.}
\label{AlphaQ_results}
\end{figure*}
To compute the speedup and efficiency, we redistributed the data at each step in order to vary the computing resources used. 
The same domain partitioning strategy implemented by the simulation is used: an equal volume decomposition, thus, the sub-volumes become smaller as the number of processes increases. 
The effect of this is seen as less of a speedup and lower parallel efficiency when using more processes, see Figure~\ref{AlphaQ_results}. 
The issue is related to worsening performance at later simulation time when entering the highly clustered regime. 
When the domain becomes highly clustered, small sub-volumes create large data imbalances. 
For example with 256 processes at $z \approx 1.0$, the volume with the most data had 4,114,405 particles, with the mean number of 1,701,349 particles across all sub-volumes, resulting in a 83\% relative difference. 
In contrast, with 64 processes at $z \approx 7.31$, the relative difference was only 39\%. 
This leads to the question of trying to load balance the data to achieve better speedup and efficiency.
Looking at Figure~\ref{breakdown}, it is not clear what may be the best strategy. 
There is not enough computation performed to warrant the costly data movement required for load balancing using the current strategy.  
Lastly, we examine the memory overhead of executing our code using the system function \texttt{getrusage}. 
We log the maximum memory utilization of the compute nodes, each running 16 MPI processes. 
\begin{table}[h!]
\centering
\label{my-label}
\begin{tabular}{llll}
$N_{ranks}$ & $<N_p>$ & $<N_h>$ & GB \\
256 & 1,797,569 & 15,283 & 3.3 \\
128 & 3,595,138 & 30,566 & 7.6 \\
64  &  7,190,276 & 61,132 & 10.9 \\
32  & 14,380,552 & 122,265 & 21.25
\end{tabular}
\end{table}

\begin{figure}[h!]
\begin{center}
\includegraphics[scale=.4]{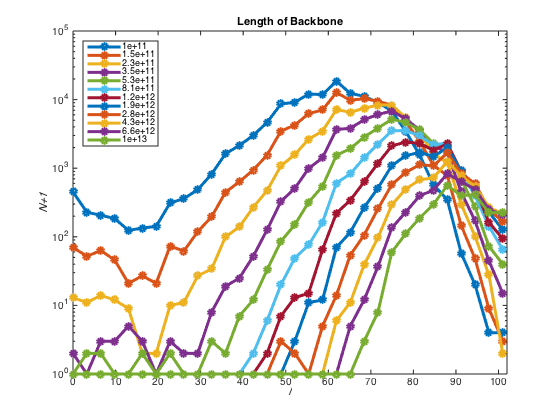}
\end{center}
\caption{The length of the merger tree backbone, i.e., the path of the most massive progenitor, for halos in the final timestep. The length gives a measure of how rubustly single halos can be tracked through the complicated interactions of structure formation.}
\label{bb_len}
\end{figure}

There exist several commonly accepted measures to characterize the quality, or correctness, of merger trees~\cite{srisawat2013sussing}. One such measure is the length of the merger tree backbone, which is the path of the most massive progenitor, see Figure~\ref{bb_len}. We partition the halos into narrow mass bins and plot the distribution of the backbone lengths in each bin to generate the figure. 
Our results show a robust identification of the formation history, consistent with what has been seen by other merger tree codes on cosmological simulations. Intuitively, the results show that larger, high-mass halos tend to have a longer formation history than lower mass halos. 

\begin{figure}[tbp]
\begin{center}
\includegraphics[scale=.6]{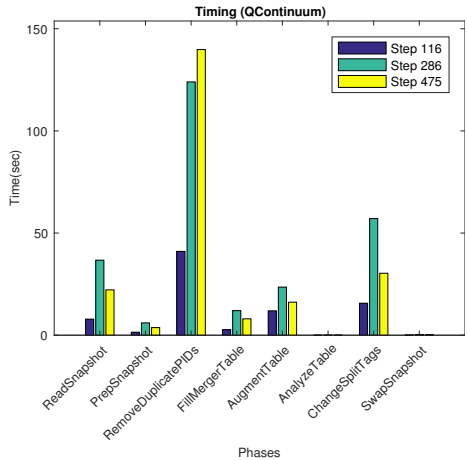}
\end{center}
\caption{The timing breakdown for the individual phases of building merger trees for the Q Continuum simulation. }
\label{qc_time}
\end{figure}

To test our implementation on an extreme scale simulation we ran our code on the complete 101 halo analysis snapshots for Q Continuum simulation using 16,348 processes. 
Each halo analysis snapshot of the simulation has a halo-particle membership function that maps approximately half of the simulation particles to halos, that is, the halo finder determined that about half of the particles belong to halos, with the remaining particles being field particles. 
The Q Continuum simulation was run with a 3D volume decomposition of $32\times32\times16$, resulting in an average of 14,797,466 particles per process, which is roughly equivalent to the AlphaQ simulation with 32 processes with a $4\times4\times2$ decomposition. 
We see a memory utilization maximum of 17.5 GB when running with 8 ranks/node due to the increased load imbalance.
In Fig \ref{qc_time}, we show the timing breakdown for building merger trees for selected snapshots of the Q Continuum simulation. 
Step 286 is most similar to the timing breakdown with 32 processes of the AlphaQ simulation shown in Fig \ref{breakdown}, both in the average number of particles per process and in simulation time. 
Although we do still see similar relative execution times for the phases, e.g., \texttt{RemoveDuplicates} and \texttt{ChangeSplitTags} far outweigh others, the cumulative end-to-end execution is greater with Q Continuum. 
The explanation is largely due to an even greater load imbalance stemming from the higher mass resolution of the Q Continuum simulation, which is approximately an order of magnitude greater than AlphaQ. 
The Q Continuum simulation covers a volume of (1300Mpc)$^3$ with a decomposition of $32\times32\times16$, making each sub-volume approximately ($40\times40\times80$)Mpc$^3$. 
As we discussed, when holding the number of particles per process fixed, the resulting AlphaQ simulation that covers a volume of (360Mpc)$^3$ has a decomposition of $4\times4\times2$, which results in sub-volumes of ($90\times90\times180$)Mpc$^3$. 
With the equal volume decomposition, the load imbalance is attributed to the clustering that naturally occurs in cosmological simulations where the homogeneity scale is closer to 200 Mpc. 

\begin{figure}[h!]
\begin{center}
\includegraphics[scale=.2]{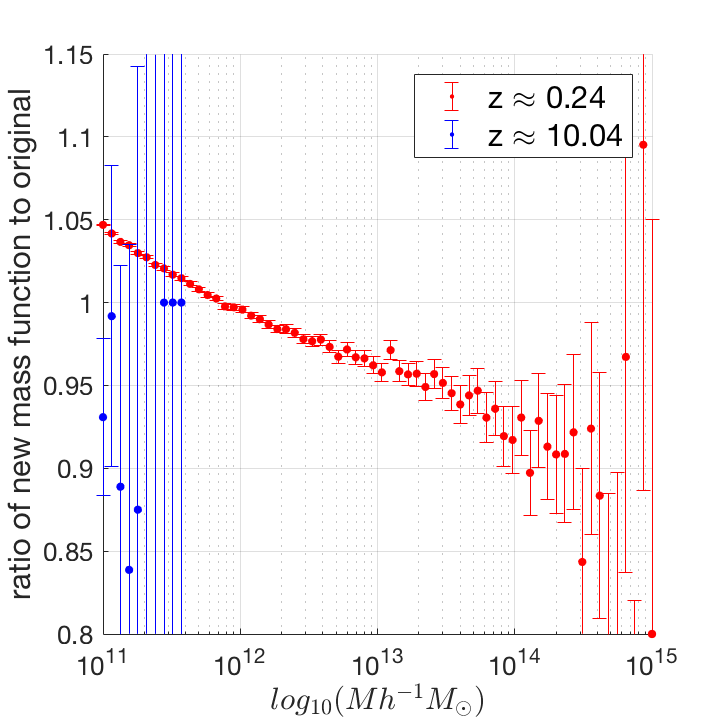}
\end{center}
\caption{The ratio of the new halo mass function obtained after applying our merger tree splitting algorithm to the original; taken from the Q Continuum Simulation at a very early and late simulation time. The effect of splitting is clearly visible with more small halos being introduced and in turn having fewer large halos. The result is comparable to that obtained by a subhalo splitting algorithm ~\cite{genel2009halo}.}
\label{mf_ratio}
\end{figure}

As a qualitative test we compared the mass function ratio in Figure \ref{mf_ratio} to  results obtained from a halo splitting algorithm using subhalo merger trees~\cite{genel2009halo}. 
We see a very similar trend and although a direct comparison is not possible for several reasons, e.g., a different FOF linking length, the results look consistent and certainly produce more trees with more physically consistent properties.

\section{Conclusion}
We have presented a novel method for building merger trees by splitting non-hierarchical FOF halos and for using the cores of merging halos to robustly track the evolution of substructure. This method presents a number of advantages over conventional techniques. It is scalable to the largest state-of-the-art simulations and does not require complex subhalo definitions and tracking, as other methods do. The core tracking methodology allows for substructure information to be included in merger tree construction in a way that avoids the subhalo disruption problem, and indeed even the identification of subhalos. In applications to galaxy formation modeling, the subhalo mass is typically not used, but rather the mass of the halo just before it accretes onto the major halo, i.e., the `infall' mass. Since the infall mass is present in our approach, there is no loss of generality when applying it to model galaxies in the construction of synthetic catalogs.


\section*{Acknowledgment}
Argonne National Laboratory's work was supported under U.S. DOE contract DE-AC02-06CH11357. 
Partial support was provided by the SciDAC program funded by the U.S. DOE, Office of Science, jointly by Advanced Scientific Computing Research and High Energy Physics and by an Argonne LDRD award. 
This research used resources of the Argonne Leadership Computing Facility, supported by DOE/SC under contract DE-AC02-06CH11357, and of the Oak Ridge Leadership Computing Facility, supported by DOE/SC under contract DE-AC05-00OR225.
This work is supported in part by the following grants: NSF award CCF-1409601; DOE awards DE-SC0007456, DE-SC0014330.
\bibliographystyle{IEEEtran}
\bibliography{./bibtex}
\end{document}